# Physics-Informed Regression Modelling for Vertical Façade Surface Temperature: A Tropical Case Study on Solar-reflective Material

Shisheng Chen[a, b], Shanshan Tong[b, *], Nyuk Hien Wong[b], May Lwin Oo[b], Joie Lim[b], Erna Tan[b], Ruohan Xu[b], Marcel Ignatius[b], Yang He[c] a-School of Architecture and Urban-Rural Planning, Fuzhou University, Fuzhou, PR China, 350108.

b-Department of the Built and Environment, National University of Singapore, 4 Architecture Drive, Singapore, 117566.

c-College of Environmental Science and Engineering, Donghua University, Shanghai, PR China, 201620.

* Corresponding author.

E-mail address: stong@nus.edu.sg (S Tong)

**Abstract**:

Urban heat islands (UHI) pose a critical challenge in densely populated cities and tropical climates where large amounts of energy are used to meet the cooling demand. To address this, Building and Construction Authority of Singapore provides incentives for passive cooling such as using of solar-reflective material in its Green Mark guidelines. Thus, understanding about its real-world effectiveness in tropical urban environments is required. Previous field evaluations of solar-reflective material have been limited by short monitoring periods, applications to horizontal surfaces in tropical regions, and the absence of physics-constrained regression tools for predicting vertical façade temperatures. This study addressed these gaps using a hybrid modelling framework combining a transient physical model and data driven model through year-long field measurements. Several machine learning algorithms were compared including multiple-linear regression (MLR), random forest regressor (RF), AdaBoost regressor (AB), extreme gradient boosting regressor (XGB), and TabPFN regressor (TPR). The results indicated that the transient physical model overestimated façade temperatures in the lower temperature ranges. The physics-informed MLR achieved best performance with improved accuracy for pre-cool paint ($R^2$=0.96, RMSE=0.83°C) and post-cool paint ($R^2$=0.95, RMSE=0.65°C) scenarios, reducing RMSE by 26% and 44%, respectively. The hybrid model also effectively predicted hourly heat fluxes revealing substantial reductions in surface temperature and heat storage with increasing albedo. The maximum net heat flux $q_{net}$ was reduced by about 30-65 W/m² in the post-cool paint stage (albedo = 0.73) compared to the pre-cool paint stage (albedo = 0.31). As albedo increases from 0.1 to 0.9, the sensitivity analysis predicts that the maximum daytime surface temperature will decrease by about 11°C and the peak heat release of the net heat flux will decrease significantly from about 161 W/m² to 27 W/m². These findings underscore the effectiveness of physics-informed regression model for vertical façade surface temperature prediction and value of high-albedo surface in minimizing heat absorption and thermal exchange.

**Key words**: Physics-Informed Regression Model; Vertical Facade Surface Temperature; Solar-reflective; Passive Cooling;

## 1. Introduction

Urban heat islands (UHI) represent a critical environmental challenge in contemporary cities and demand urgent mitigation efforts. The UHI phenomenon is primarily attributed to anthropogenic heat emissions from energy use, the thermal storage characteristics of construction materials, reduced evaporative cooling caused by the loss of vegetation and water bodies, and heat trapping resulting from high-density urban morphology (Building and Construction Authority, 2021). In tropical city-states such as Singapore, which experiences persistently hot and humid conditions, the extensive reliance on air-conditioning systems leads to substantial building energy consumption, placing considerable pressure on both urban management authorities and building occupants. In response, the Building and Construction Authority (BCA) of Singapore has introduced green building design standards aimed at enhancing energy efficiency. Among various passive cooling measures, cool paints with a high solar reflectance index have been identified as one of the most effective solutions for the local climate and are encouraged under the BCA Green Mark 2021 scheme, particularly for applications covering more than 50% of a site area (Building and Construction Authority, 2021).

Radiative cooling (RC) is a passive cooling approach that utilizes the atmospheric transparency window (8–13 μm) to release thermal energy into outer space, offering a sustainable pathway for reducing building cooling demand and mitigating urban heat accumulation (Chen et al., 2020; Li et al., 2020). With ongoing developments in material science, RC technologies have diversified and can be categorized according to their functional roles, including selective emitters, reflective layers, back-reflector systems, matrix and insulation materials, as well as dynamic switching components (Li et al., 2020). Contemporary RC systems are capable of providing continuous cooling throughout the day and night through two fundamental mechanisms: first, high infrared emissivity that enhances longwave thermal radiation emission (Yan et al., 2024); and second, elevated solar reflectivity that minimizes solar heat absorption (Li et al., 2020). High surface albedo reduces daytime radiative heat gain by reflecting incoming solar radiation, while strong emissive properties promote nocturnal heat dissipation, thereby lowering cooling demand across diurnal cycles.

Empirical studies have demonstrated that the integration of RC technologies in buildings can yield considerable energy and economic benefits. Field investigations conducted in tropical and subtropical regions reported cooling energy savings ranging from 20% to 40% over a 20-year building lifespan, accompanied by reductions in global warming potential due to decreased $CO_2$ emissions associated with energy use (Pirvaram et al., 2022). Research examining super-cool roof systems found substantial decreases in roof heat transfer values, with heat gain reductions between 73.9% and 90.7%, resulting in notable energy and cost savings (Chen et al., 2024). In another study, the installation of RC films on a large-scale commercial warehouse led to a 44.36% reduction in daily cooling electricity demand compared with conventional steel roofing, while calibrated simulations predicted annual cooling energy savings of up to 65.2% across various hot climate regions (Wang et al., 2022). Additionally, physics-based modelling suggested that RC technologies could reduce cooling energy consumption by approximately 25%, delivering average cooling power levels of 43–45 $W/m^2$ in tropical climates (Hanif et al., 2014). At the urban scale, simulations indicated that RC-treated surfaces could lower nighttime temperatures by 1.5 ℃ and daytime temperatures by 7.7 ℃ relative to conventional concrete roofs, with peak temperature differences reaching 15.5 ℃ at midday (Li et al., 2023). This study also reported carbon emission reductions of 0.52 $gCO_2/(m^2 \cdot h)$ for roofs and 0.16 $gCO_2/(m^2 \cdot h)$ for walls. Despite extensive research on cool roof applications, investigations into the performance of RC materials on vertical building façades remain limited. Full-scale experimental studies on prefabricated residential buildings in subtropical climates revealed that RC coatings applied to roofs significantly outperformed those on façades (Wang et al., 2025). Under ambient conditions, roof surfaces exhibited 132.96% greater daytime cooling and 75.28% greater nighttime cooling compared with white putty, whereas façades achieved only 13.30% cooling during the day and experienced negative performance (−47.5%) at night. When air-conditioning systems were operating, roof cooling advantages persisted, with improvements of 55.51% during daytime and 28.31% at night, while façade performance gains were marginal at 3.58% during the day and substantially negative (−139.13%) at night. These findings underscore the critical role of view factors in RC effectiveness, indicating that vertical façades possess significantly lower cooling potential than horizontal roof surfaces.

The evaluation of building thermal behavior typically relies on either physics-based numerical simulations or data-driven modelling techniques. While physics-based methods provide strong theoretical foundations, they are computationally intensive when applied to complex urban settings involving multi-scale and coupled heat transfer processes. Data-driven approaches, by contrast, offer efficient pattern recognition capabilities but suffer from limited generalization, high dependence on large datasets, and poor interpretability due to their black-box nature. Integrating these two modelling paradigms presents an opportunity to overcome their individual limitations and deliver robust, site-specific solutions for building performance analysis.

Physics-informed machine learning (PIML) has recently emerged as an effective framework for addressing complex scientific and engineering challenges governed by physical laws and differential equations, combining the strengths of machine learning with physics-based constraints (Karniadakis et al., 2021). By embedding physical principles directly into neural network architectures through physics-informed loss functions or hybrid modelling strategies, PIML achieves improved accuracy and generalization even when training data are sparse or noisy. This methodology has demonstrated strong performance in a wide range of applications, including computational fluid dynamics, where it has been shown to solve Navier–Stokes equations more efficiently than conventional numerical solvers (Sharma et al., 2023; Wang et al., 2020). In the building domain, PIML approaches have been proposed for optimizing HVAC system operations by integrating physical simulations with data-driven models, enabling improved control of thermal comfort, indoor air quality, and energy consumption (Bünning et al., 2022; Chen et al., 2023; Gokhale et al., 2022; Ma et al., 2024; Wang & Dong, 2023). Conventional building thermal modelling typically relies on either white-box physical models or black-box data-driven techniques. Tools such as EnergyPlus and CFD offer high-fidelity simulations but require extensive input parameters and substantial computational resources, making them difficult to calibrate and impractical for large-scale analyses. Data-driven models, including neural networks, provide computational efficiency and flexibility but are vulnerable to overfitting and may generate predictions that violate physical laws. Hybrid modelling frameworks have therefore become increasingly important for urban-scale environmental applications, including land surface temperature simulation for heat island analysis (Chen et al., 2022), urban flood prediction (Yang et al., 2024), and enhanced weather and climate forecasting with improved spatiotemporal generalization (Kashinath et al., 2021). However, despite their growing adoption, such approaches have not yet been applied to predicting vertical façade surface temperatures, a key variable in assessing surface thermal behavior and building energy performance.

This study examined the implementation of solar-reflective radiative cooling materials on an institutional building in Singapore, selected due to its intensive daytime air-conditioning usage and the building owner's interest in reducing peak cooling demand. Vertical façade surface temperatures were monitored using thermocouples and infrared imaging, while local environmental conditions were recorded via a mobile weather station. The thermal performance of solar-reflective materials on vertical façades is governed by coupled radiative, conductive, and convective heat transfer processes in accordance with thermodynamic principles. Although simulation platforms such as EnergyPlus and CFD are capable of accurately modelling façade thermal behavior, they require comprehensive input datasets—including detailed boundary conditions, three-dimensional building geometries, material properties, and significant computational effort—rendering them unsuitable for large-scale or long-term applications. Conversely, solar irradiance simulations on building façades are relatively straightforward and computationally efficient; however, converting irradiance outputs into reliable façade temperature predictions remains challenging due to the complex interactions among heat transfer mechanisms. To address this limitation, the present study developed a PIML framework that links solar radiation inputs with façade thermal responses, maintaining physical consistency while significantly reducing computational demands.

It was hypothesized that transient physics-based models could provide an initial approximation of façade surface temperatures, while regression-based techniques informed by field measurements could correct systematic deviations and align predictions with observed data. By integrating these complementary data sources, physics-informed regression models (PIRM) were established to improve the accuracy and reliability of vertical façade temperature predictions.

## 2. Methodology

### 2.1 Research Design

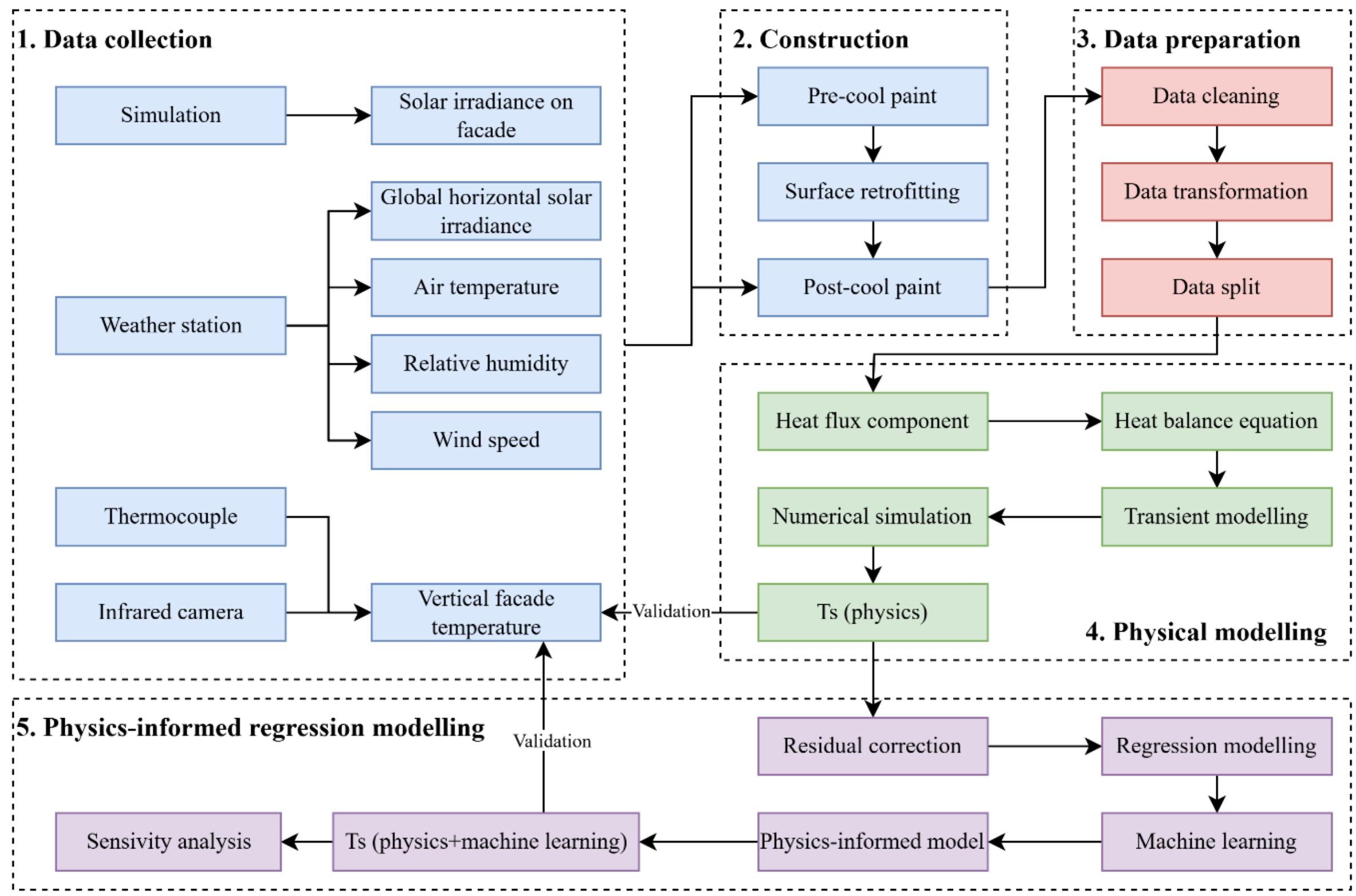

Figure 1: Research design

The overall research framework, as presented in Figure 1, consisted of five sequential phases: data acquisition, construction, data processing, physical modelling, and physics-informed regression modelling. The study began with the systematic collection of continuous datasets obtained before and after the application of cool paint on the building façade. Solar radiation incident on the vertical façade was simulated using the Ladybug plugin within the Grasshopper environment. Concurrently, a mobile weather station was deployed to record on-site environmental parameters, including global horizontal irradiance, ambient air temperature, relative humidity, and wind speed. Surface temperatures of the vertical façade were monitored through the installation of thermocouples and the use of an infrared camera to obtain thermal imagery. Together, these measurements provided a comprehensive dataset for characterizing the environmental and thermal conditions affecting façade performance.

After completing the data collection and construction stages, the raw datasets were processed to ensure their suitability for subsequent analyses. This stage involved data screening and cleaning procedures to eliminate noise, outliers, and irrelevant records, thereby improving data reliability. The processed data were then transformed into a structured tabular format compatible with modelling workflows. For model development and evaluation, the dataset was divided into training and validation subsets.

The next phase focused on analysing heat flux components to characterize heat transfer across the façade surface. A heat balance formulation was employed to describe the overall energy exchange within the system, while transient analysis was used to capture the temporal evolution of thermal behaviour. Numerical simulation methods were applied to represent the thermal response of the painted façade surfaces, and the simulation outputs were validated against experimental measurements. This phase was essential for elucidating the underlying physical mechanisms governing heat transfer on vertical building façades.

To further improve prediction accuracy, regression modelling and machine learning approaches were introduced to complement the physical models. A residual correction strategy was implemented to account for systematic differences between simulated and measured façade temperatures. An additional validation step was conducted to assess the consistency between physics-informed modelling results and experimental observations. Moreover, a sensitivity analysis was performed to evaluate the influence of surface albedo on vertical façade temperature and to investigate the thermal response of solar-reflective cool paint. In this analysis, hourly-averaged values of measured air temperature, relative humidity, wind speed, and simulated incident solar radiation on the vertical façade were adopted, as discussed in Section 3.3.

## 3. Results

### 3.1 Ambient weather conditions

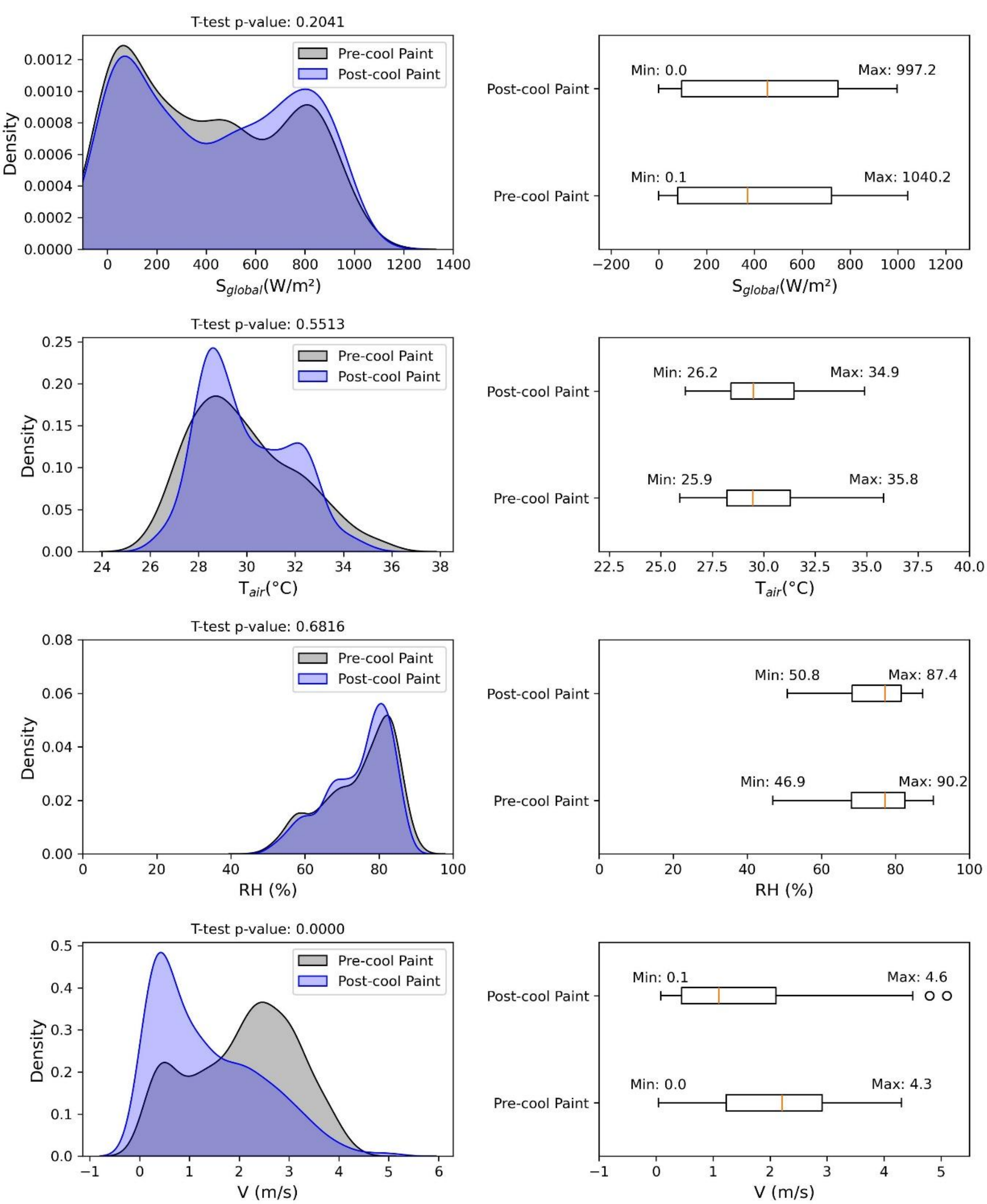

Figure 2: Weather profile during the experiment period

Table 1: Environmental conditions

| Statistics | $S_{global}$ | | $Tair$ | | $RH$ | | $V$ | |
|---|---|---|---|---|---|---|---|---|
| | Pre-cool Paint | Post-cool Paint | Pre-cool Paint | Post-cool Paint | Pre-cool Paint | Post-cool Paint | Pre-cool Paint | Post-cool Paint |

| mean | 408.9 | 437.8 | 29.8 | 29.9 | 74.6 | 74.4 | 2.1 | 1.4 |
|---|---|---|---|---|---|---|---|---|
| std | 319.8 | 325.1 | 2.2 | 1.9 | 9.5 | 8.7 | 1.1 | 1.1 |
| min | 0.1 | 0.0 | 25.9 | 26.2 | 46.9 | 50.8 | 0.0 | 0.1 |
| 25% | 78.7 | 96.0 | 28.2 | 28.4 | 68.1 | 68.4 | 1.2 | 0.4 |
| 50% | 370.8 | 454.7 | 29.5 | 29.5 | 77.2 | 77.2 | 2.2 | 1.1 |
| 75% | 723.6 | 749.2 | 31.3 | 31.5 | 82.5 | 81.5 | 2.9 | 2.1 |
| max | 1040.2 | 997.2 | 35.8 | 34.9 | 90.2 | 87.4 | 4.3 | 5.1 |

### 3.2 Comparisons of vertical surface temperature predictions

Table 2: Model performance for predicting vertical façade temperature

| Phase | Model | Training RMSE | Training $R^2$ | Test RMSE | Test $R^2$ |
|---|---|---|---|---|---|
| Pre-cool paint | Physics | N/A | N/A | 1.12 | 0.92 |
| | Physics + MLR | 0.60 | 0.96 | 0.83 | 0.96 |
| | Physics + RF | 0.37 | 0.98 | 0.86 | 0.95 |
| | Physics + AB | 0.48 | 0.97 | 0.85 | 0.95 |
| | Physics + XGB | 0.38 | 0.98 | 0.85 | 0.95 |
| | Physics + TPR | 0.33 | 0.99 | 0.86 | 0.95 |
| Post-cool paint | Physics | N/A | N/A | 1.17 | 0.84 |
| | Physics + MLR | 0.80 | 0.96 | 0.65 | 0.95 |
| | Physics + RF | 0.61 | 0.98 | 0.68 | 0.95 |
| | Physics + AB | 0.67 | 0.97 | 0.69 | 0.94 |
| | Physics + XGB | 0.64 | 0.97 | 0.66 | 0.95 |
| | Physics + TPR | 0.39 | 0.99 | 0.77 | 0.93 |

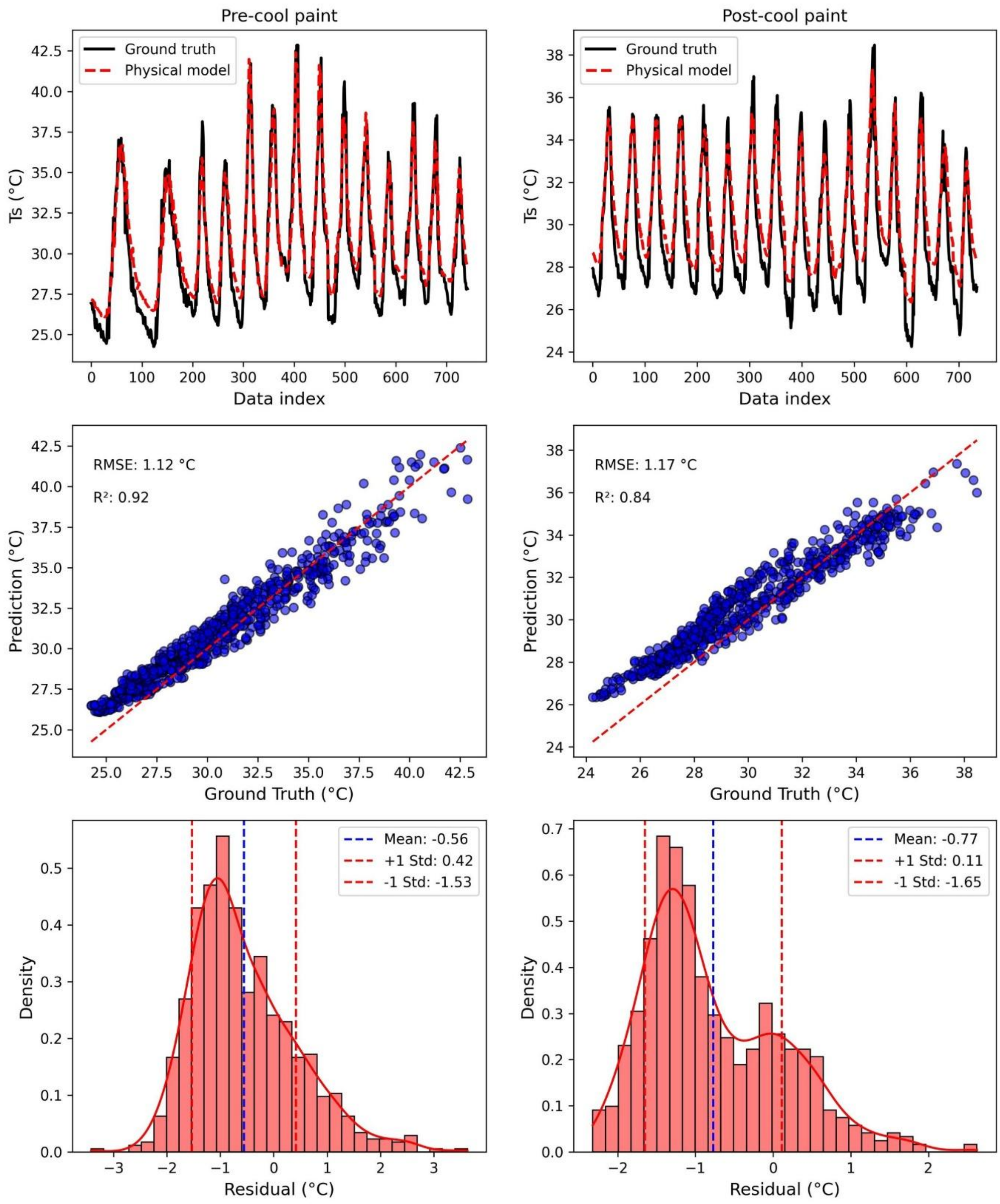

Figure 3: Predicted vertical façade surface temperature by transient physical model

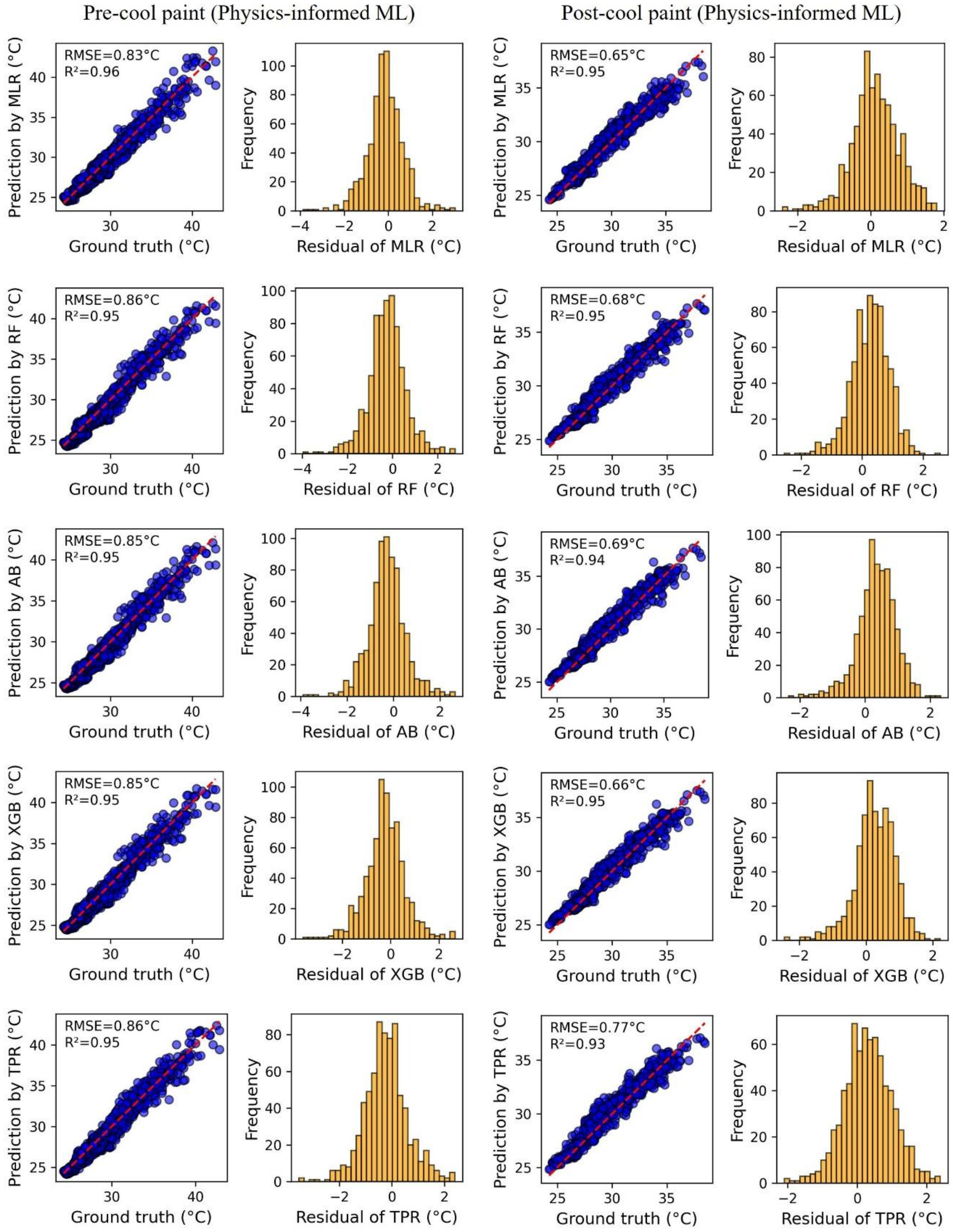

Figure 4: Comparison of prediction performance among PIRM

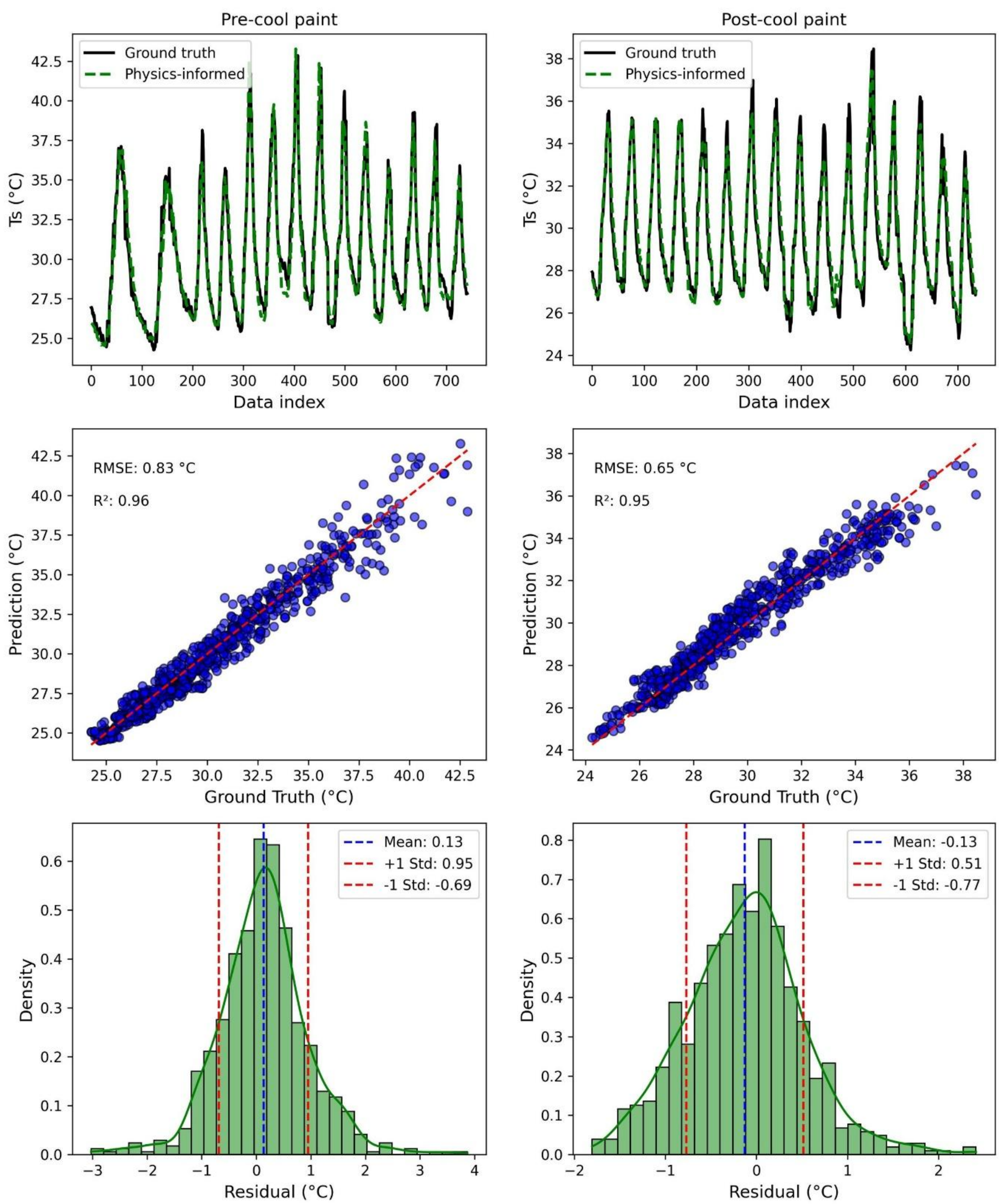

Figure 5: Predicted vertical façade surface temperature by physics-informed regression model (MLR)

### 3.3 Comparisons of heat flux predictions

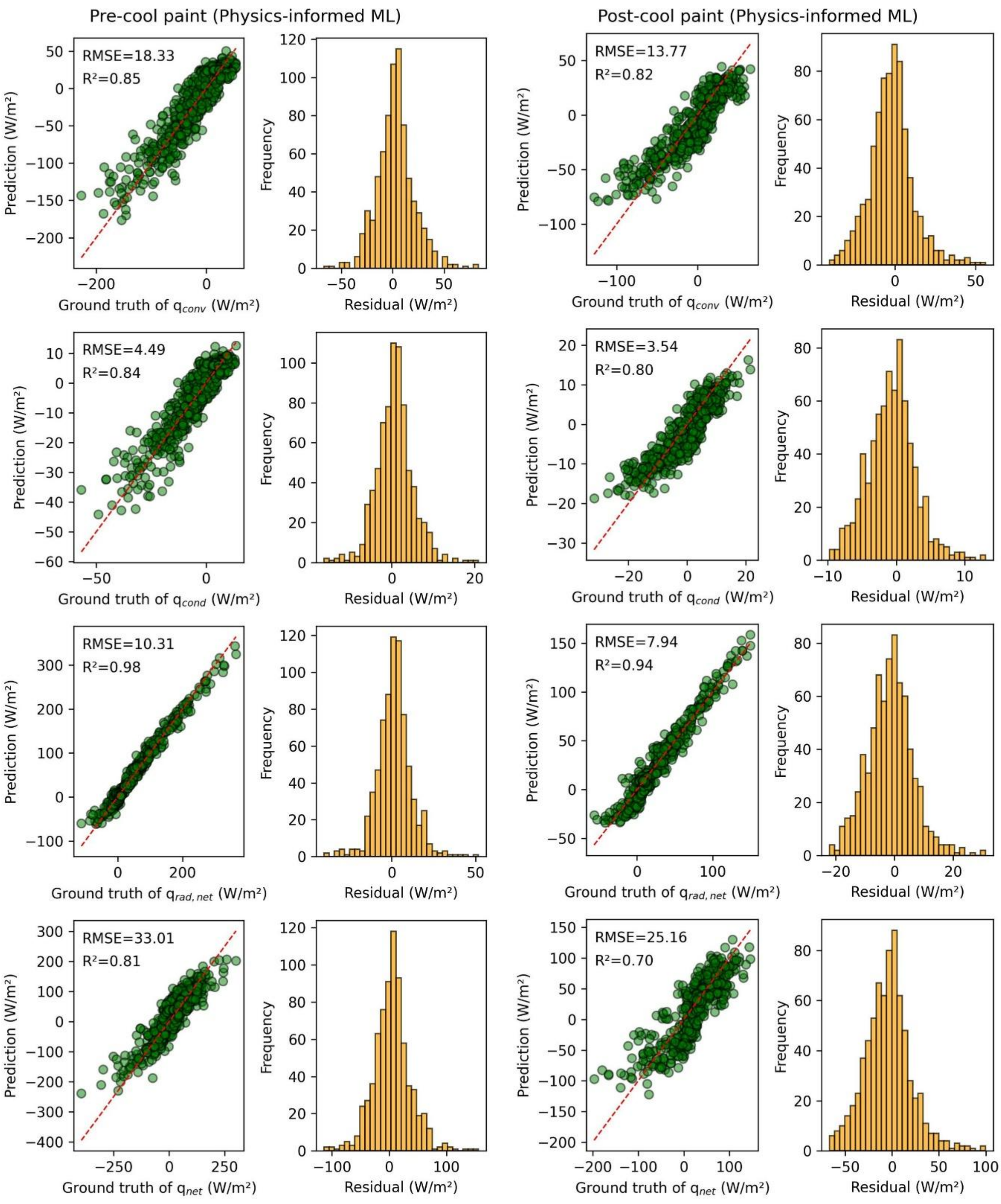

Figure 6: Heat flux estimated by the physics-informed regression model (MLR) using unaggregated time series data

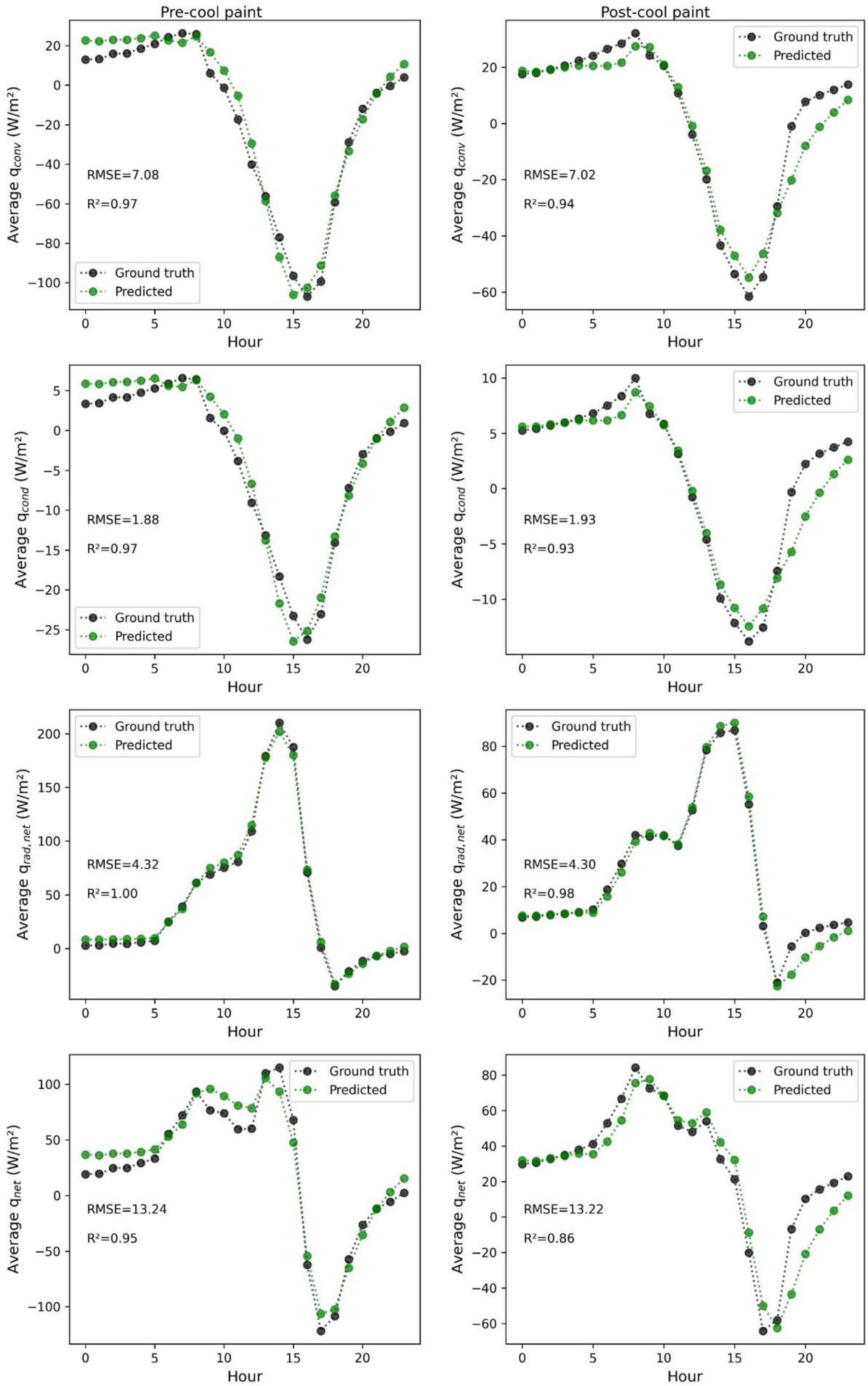

Figure 7: Hourly average heat flux estimated by the physics-informed regression model (MLR)

### 3.3 Sensitivity analysis of albedo on vertical façade surface temperature

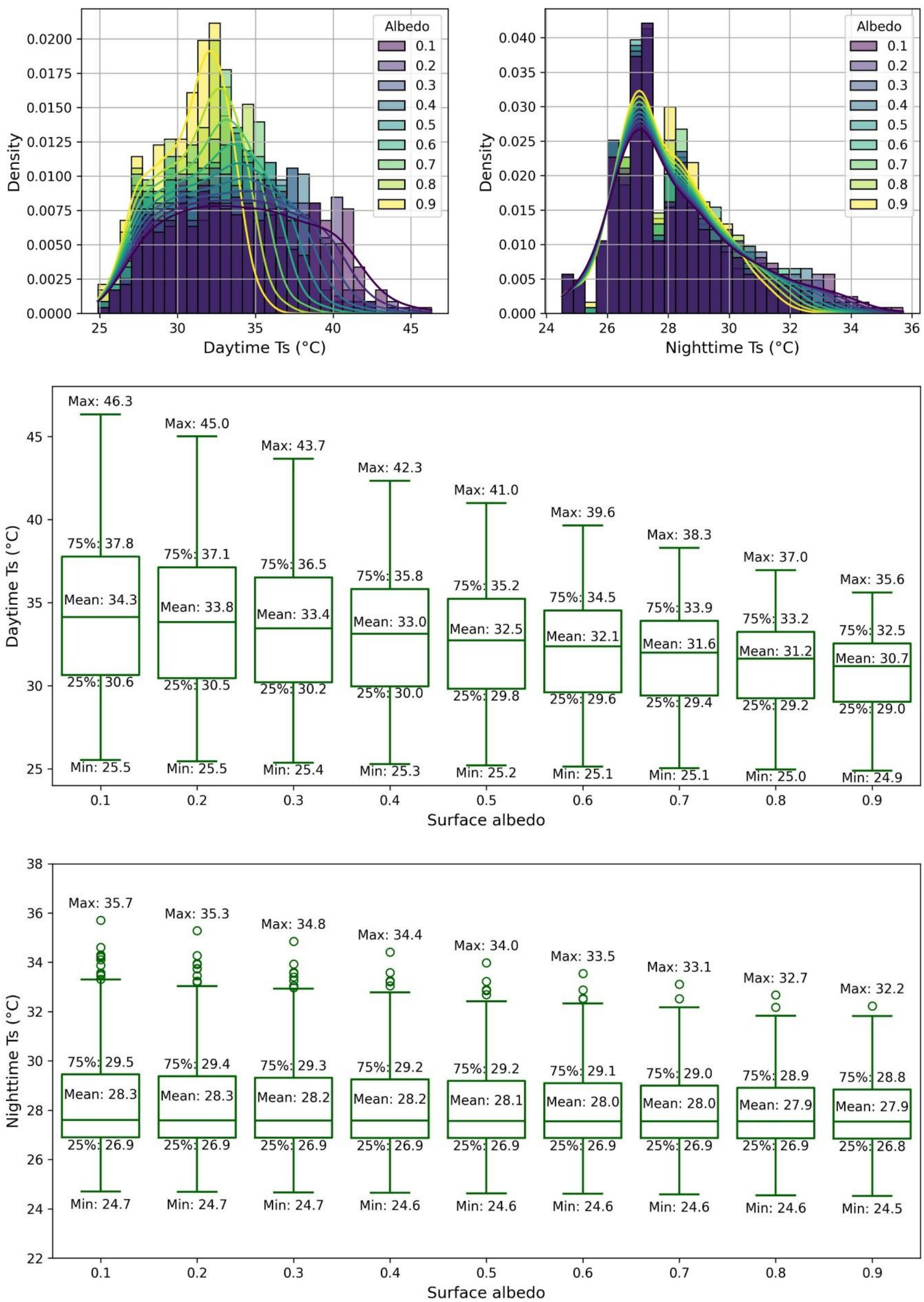

Figure 8: The impact of albedo on vertical façade surface temperature at different periods

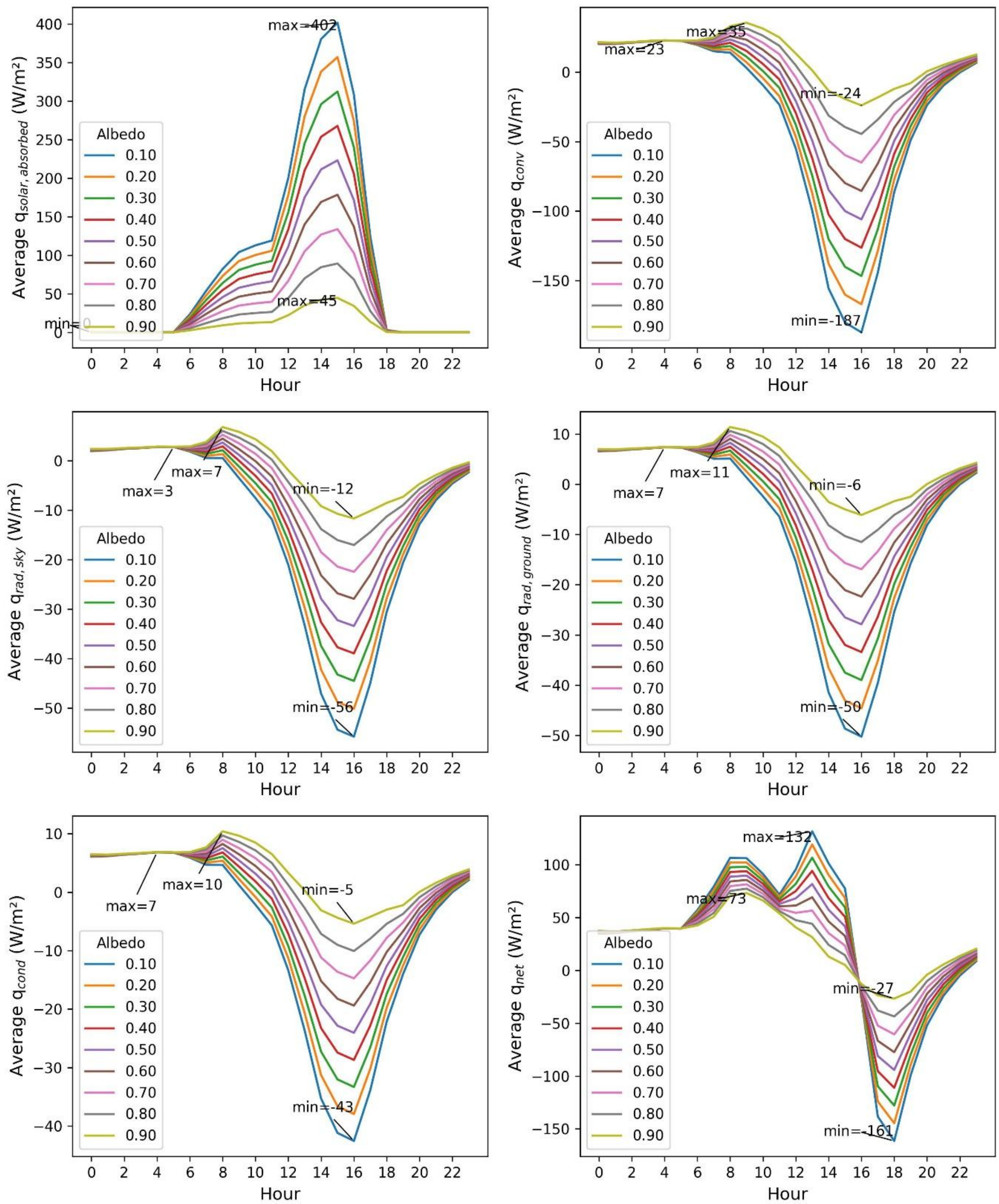

Figure 9: The impact of albedo on heat flux components at different hours

## 5. Conclusions

The standalone transient physical model showed a tendency to overpredict vertical façade surface temperatures, particularly within the lower temperature range, under both pre-cool paint ($R^2$ = 0.92, RMSE = 1.12 °C) and post-cool paint ($R^2$ = 0.84, RMSE = 1.17 °C) conditions. In contrast, the physics-informed multiple linear regression (MLR) model achieved substantially improved predictive performance for façade surface temperature in both scenarios, with $R^2$ values of 0.96 and 0.95 and RMSE values of 0.83 °C and 0.65 °C for pre-cool and post-cool paint cases, respectively. The physics-informed MLR approach also outperformed other machine learning models, including RF, AB, XGB, and TPR. Relative to the standalone transient physical model, the integration of physics-informed regression reduced the prediction error by 26% for the pre-cool paint case and by 44% for the post-cool paint case.

The MLR-based physics-informed regression model further demonstrated strong capability in reproducing hourly averaged heat flux components for both surface conditions, with coefficients of determination ranging from 0.86 to 1.0. For the net heat flux, $q_{net}$ exhibited $R^2$ values between 0.86 and 0.95, with RMSE values of approximately 13.22–13.24 W/m². Under pre-cool paint conditions, $q_{net}$ varied from approximately −130 to 115 W/m², while for post-cool paint surfaces the range was reduced to about −65 to 85 W/m². The application of solar-reflective materials led to a reduction in peak $q_{net}$ on the order of 30–65 W/m². Temporal analysis indicated that maximum heat gain occurred around hour 13, whereas peak heat loss was observed near hour 17. The net radiative heat flux, $q_{(rad,net)}$, remained predominantly positive, with the highest heat absorption occurring between hours 14 and 15. In contrast, both convective ($q_{conv}$) and conductive ($q_{cond}$) heat fluxes were largely negative, reaching their maximum magnitudes around hour 16.

Sensitivity analysis revealed a pronounced decrease in surface temperature and corresponding heat fluxes with increasing surface albedo. When albedo increased from 0.1 to 0.9, the 75th percentile of daytime surface temperature declined from 37.8 °C to 32.5 °C, while the maximum daytime temperature decreased from 46.3 °C to 35.6 °C. Correspondingly, absorbed solar heat flux, $q_{(solar,absorbed)}$, was reduced substantially from approximately 402 W/m² to 45 W/m². The net heat release represented by $q_{net}$ also decreased markedly, dropping from around 161 W/m² to 27 W/m². These findings demonstrate the effectiveness of solar-reflective materials in regulating façade surface temperatures, thereby contributing to improved building energy performance and enhanced indoor thermal comfort.

**Conflicts of Interest**: The authors declare no conflict of interest.

**Acknowledgements**

This research was supported by: (1) Supporting Cooling NUS with Baseline-Evaluating-Action-Monitoring (BEAM) initiative from National University of Singapore; (2) The Fujian Province Young and Middle-aged Teacher Education and Research Project Funding (JAT241009) (3) The Open Fund of Key Laboratory of Ecology and Energy Saving Research of Dense Habitat, Ministry of Education (20240105); (4) Fuzhou University Research Starting Fund (511470); (5) Fuzhou University Testing Fund of Precious Apparatus (2025T031).